# Title: Fragmenting Diffusion Pathways Confers Extraordinary Radiation Resistance in Refractory Multicomponent Alloys

**Authors:** Bin Xing[1,†], Bijun Xie[2,†], Wanjuan Zou[2], Eric Lang[3], Evgeniy Boltynjuk[4], Hangman Chen[2], Michael P Short[5], George Tynan[6], Timothy J Rupert[7], Jason Trelewicz[8], Horst Hahn[4,9], Blas P Uberuaga[10], Khalid Hattar[11], Penghui Cao[1,2,*]

**Affiliations:**
[1]Department of Material Science and Engineering, University of California, Irvine, California, USA
[2]Department of Mechanical and Aerospace Engineering, University of California, Irvine, CA, USA
[3]Department of Nuclear Engineering, University of New Mexico, Albuquerque, NM, USA
[4]Institute of Nanotechnology, Karlsruhe Institute of Technology, Karlsruhe, Germany
[5]Department of Nuclear Science and Engineering, Massachusetts Institute of Technology, Cambridge, MA, USA
[6]Department of Mechanical and Aerospace Engineering, University of California, San Diego, CA, USA
[7]Department of Materials Science and Engineering, Johns Hopkins University, Baltimore, MD, USA
[8]Department of Materials Science and Chemical Engineering, Stony Brook University, Stony Brook, NY, USA
[9]Department of Materials Science and Engineering, University of Arizona, Tucson, AZ, USA
[10]Material Science and Technology Division, Los Alamos National Lab, Los Alamos, NM, USA
[11]Department of Nuclear Engineering, University of Tennessee, Knoxville, TN, USA

†These authors contributed equally to this work; *Corresponding author. Email: caoph@uci.edu

**Abstract:** The accumulation and growth of vacancy clusters under irradiation is a pivotal degradation mode for structural materials in extreme environments. Even tungsten undergoes rapid defect coarsening compromising its integrity. Here we show a tungsten multicomponent alloy that effectively fragments the vacancy diffusion network, kinetically trapping defects within localized domains. This effect originates from a broad spectrum of migration barriers and substantial vacancy-jump heterogeneity, which drives the interconnectivity of diffusion paths below the percolation threshold. Starving clusters of the necessary vacancy supply, irradiation experiments and atomic-scale defect characterizations confirm negligible defect growth as radiation doses increase by four orders of magnitude. These results provide a fundamental paradigm for percolation-engineered kinetics, offering a predictive pathway for tailoring defect diffusion and discovering inherently radiation-tolerant materials.

The quest for practical fusion energy demands structural materials capable of withstanding extreme particle irradiation while retaining mechanical and thermal integrity (*1*). Tungsten, the leading candidate for plasma-facing armor, remains fundamentally limited by its response to high-energy particle irradiation. For example, even at modest irradiation doses of ~1 dpa (displacements per atom), it experiences rapid formation of dense dislocation loops and defect clusters (*2–4*). As these defects accumulate and coarsen, they drive hardening, increased fuel retention, and reduced thermal conductivity (*5–7*). Classic microstructural approaches such as introducing high-density grain boundaries (*8, 9*) or precipitates (*10*) can act as defect sinks, but these features inevitably evolve, saturate, or embrittle under sustained flux (*11*). A promising strategy would require engineering intrinsic radiation resistance within the bulk lattice, suppressing damage accumulation at its point of origin.

At the atomic level, energetic particle irradiation of crystalline solids launches displacement cascades (*12*) that produce a residual supersaturation of vacancy and interstitial defects (*13*) (*14*). Under continued radiation, this persistent defect surplus diffuses and agglomerates into stable structures such as prismatic dislocation loops and nanoscale voids (*15–17*). The progressive coalescence and growth of vacancy defects result in swelling (*18*), a critical degradation and failure mode in extreme environments. Deciphering and subsequently tailoring the atomic-scale transport processes that govern the defect growth defines the key frontier for discovering materials with inherent radiation resistance.

In pure metals like tungsten, a constant migration barrier for vacancy enables it to jump with equal probability in all crystallographic directions. This isotropic diffusion energy landscape forms a fully connected three-dimensional network of diffusion paths, allowing vacancies to perform random walks and to reach far-field clusters (Fig. 1A). If enough diffusion paths are deactivated and the fraction of accessible paths is below the percolation threshold (*19*), the interconnected network will fragment into isolated domains (Fig. 1C). This loss of connectivity could potentially impede long-range vacancy diffusion, exhausting clusters of the species needed for growth (Fig. 1C). In multicomponent alloys (*20*), the site-to-site variation in local chemical environment creates a broad spectrum of vacancy migration barriers (*21–23*), providing a versatile handle for tuning diffusion kinetics. Yet, calculating these barriers and predicting long-timescale kinetics in chemically complex environments and across large composition spaces using first-principles methods like density functional theory (DFT) is computationally prohibitive.

Here we develop a first-principles neural network kinetics (FP-NNK) model that predicts vacancy diffusion across vast configurational space with near-DFT accuracy. We reveal that diffusion paths (Fig. S1) are effectively fragmented in a tungsten alloy WMoTa, where vacancy diffusion is kinetically confined. This kinetic trapping arises from a pronounced variation in vacancy migration barriers that reduces the network connectivity. Irradiation experiments, coupled with atomic-scale defect characterizations using advanced scanning transmission electron microscopy (STEM) and four-dimensional (4D) STEM, confirm negligible vacancy cluster growth as radiation doses increase from 0.0021 to 21 dpa, indicating its extraordinary radiation resistance.

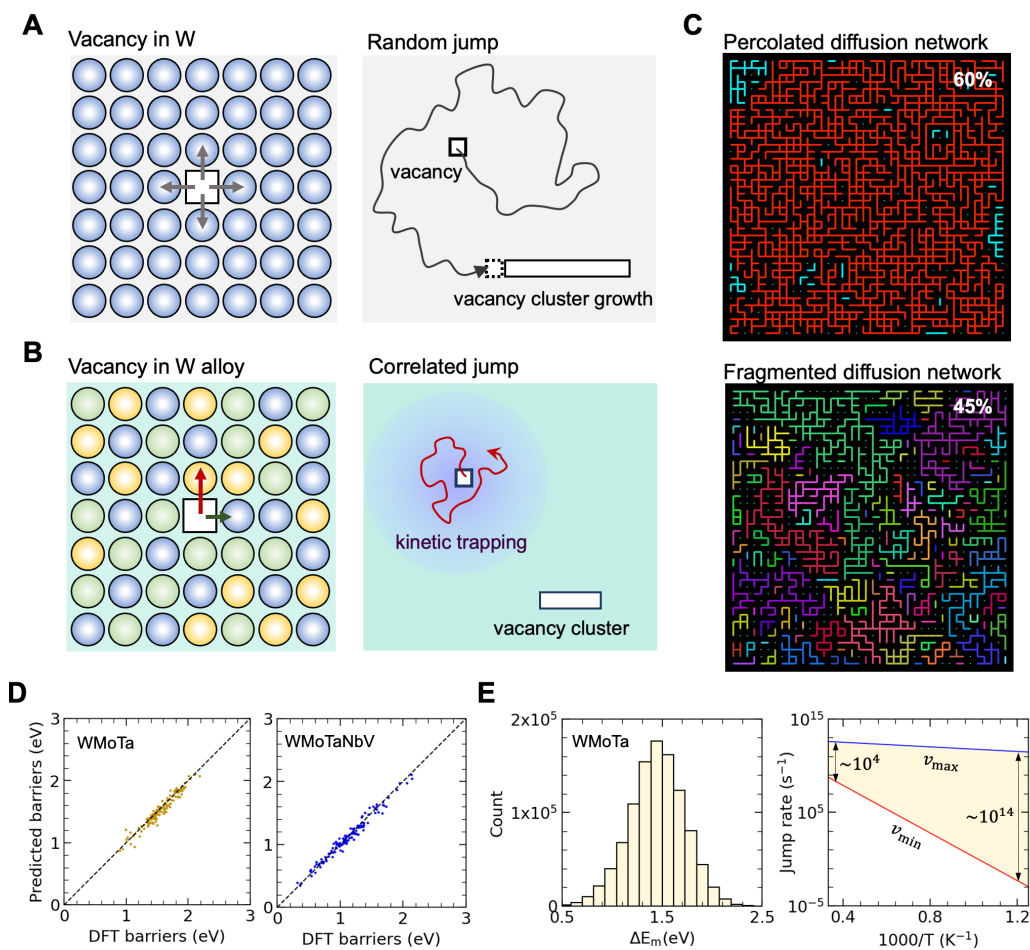

**Fig. 1. Diffusion pathways, network percolation and fragmentation, and diffusion barrier prediction in large compositional spaces.** (**A**) In pure W, a uniform migration barrier permits vacancies to jump randomly in all crystallographic directions, rendering random walks and long-range diffusion. (**B**) In alloys, local chemical fluctuations introduce heterogeneous migration barriers, leading to correlated jumps and potential confinement. (**C**) Conceptual illustration of diffusion network percolation and fragmentation. Above the percolation threshold (top, 60%), accessible pathways form a connected network, and below the threshold (bottom, 45%), the network fragments into isolated domains. (**D**) FP-NNK predictions of migration barriers compared with DFT values for ternary and quinary W alloys. (**E**) Statistical distribution of vacancy diffusion barriers in WMoTa spanning 0.5 to 2.3 eV, corresponding to heterogeneous jump rates.

## Prediction of vacancy diffusion in quinary alloying space

The FP-NNK framework integrates neuron representations of atomic structure (*24*) with DFT-calculated transition states to predict diffusion kinetics with high efficiency and precision (Fig. S2). The neuron descriptor is invariant to the number of constituent elements, important for exploring complex materials across large compositional spaces. Figure 1D benchmarks FP-NNK on the test set for WMoTa and WMoTaNbV, yielding a mean absolute error of ~0.05 eV relative to DFT, corresponding to <3.5% of the true migration barrier. Comparable performance is achieved in binary, ternary, and quaternary compositions, demonstrating its ability to predict vacancy migration barriers throughout the quinary W-Mo-Ta-Nb-V compositional space (Fig. S3). Notably, WMoTa exhibits a broad distribution of diffusion barriers ranging from ~0.5 to 2.3 eV, corresponding to jump rates that can differ by many orders of magnitude between the fastest and slowest events (Fig. 1E). To model consecutive vacancy migrations, we integrate the kinetic Monte Carlo method into the framework, efficiently molding vacancy jumps via computational efficient neuron kinetics. Because vacancy diffusion is dictated by the local chemical environment, FP-NNK trained on small-scale DFT models can be extended to simulate much larger systems with high fidelity.

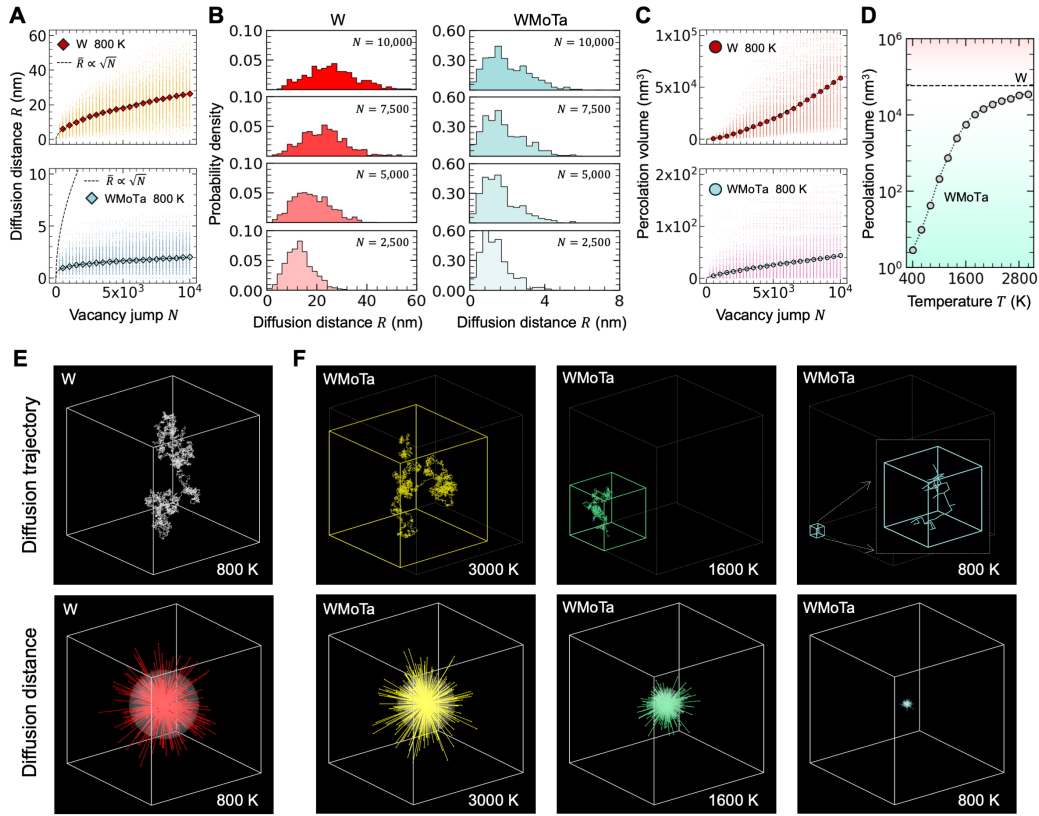

**Fig. 2. Vacancy diffusion distance, trajectories, and kinetic trapping.** (A) Mean diffusion distance $\bar{R}$ as a function of jump number $N$ for W and WMoTa at 800 K. Scattered points represent 500 independent vacancy trajectories, and filled symbols denote the average. The dashed line indicates the classical $\bar{R} \propto \sqrt{N}$ scaling. (B) Probability density distributions of diffusion distance in W (left) and WMoTa (right) after 2,500, 5,000, 7,500, and 10,000 jumps. W exhibits progressively broadening front, whereas WMoTa remains bounded at ~6 nm. (C) Evolution of percolation volume as a function of vacancy jumps. (D) Temperature dependence of the percolation volume, showing a substantial decrease in WMoTa relative to W at 400 K. (E) Representative vacancy trajectory (top) and the diffusion vectors (bottom) for 500 vacancies in W after 10,000 jumps at 800 K. (F) Corresponding trajectories and diffusion vectors in WMoTa after the same number jumps at 3000, 1600, and 800 K.

## Vacancy diffusion and kinetic trapping

To capture the stochasticity of vacancy transport, we track the diffusion trajectories of 500 independent vacancies throughout $10^4$ discrete jumps. In W, the relationship between mean diffusion distance ($\bar{R}$) and number of jumps ($N$) follows the characteristic scaling law (25), $\bar{R} \propto \sqrt{N}$, signifying random walk nature (Fig. 2A). In stark contrast, vacancies in WMoTa fundamentally deviate from the random walk behavior, and the diffusion distance exhibits an asymptotic saturation at ~2 nm, marking a breakdown of long-range transport. This localization is further evidenced by the evolution of probability density distributions (Fig. 2B). The diffusion front in W progressively broadens to ~60 nm after $10^4$ jumps. The WMoTa front, however, shows negligible expansion as $N$ increases from 5,000 to $10^4$, remaining sub-diffusive and restrained at ~6 nm. Further analysis of diffusion distance at various temperatures (Fig. S4-S5) indicates that the chemical disorder in WMoTa can restrict vacancy diffusion, preventing their expansive migration even after extensive jump sequences.

To characterize the spatial extent of this confinement, we define a percolation volume representing the minimum envelope enclosing the entire diffusion trajectory. In W, this volume expands continuously, exceeding $6 \times 10^4$ nm$^3$ after $10^4$ jumps (Fig. 2C). Conversely, the percolation volume in WMoTa exhibits only minimal expansion, below 50 nm$^3$ after the same jump number. The reduction in diffusion expansion is highly temperature-dependent, governed by the competition between thermal activation and the underlying diffusion barrier spectrum. At a high temperature of 3000 K, the percolation volume in WMoTa approaches that of pure W, indicating that thermal energy ($k_B T$) dominates over diffusion-barrier variance. As the temperature decreases to 400 K, the volume in WMoTa collapses by a few orders of magnitude (Fig. 2D). This drastic reduction signifies a transition from a regime of thermally dominated random hopping at high temperatures to one of chemical disorder-mediated kinetic trapping at low temperatures.

Visual inspection of the vacancy trajectories and individual diffusion vectors provides direct morphological evidence of the transport transition (Fig. 2, E and F). At 3000 K, the WMoTa trajectory remains spatially expansive and comparable to that of pure W, as a result of that thermal fluctuation overcoming the heterogeneous migration barriers. Upon lowering the temperature to 1600 K and 800 K, the trajectories contract into localized clusters with strongly reduced net distance. At 800 K, the vacancies are confined to small spatial regions (Fig. 2F), suggesting the emergence of local domains in the system that constrain vacancies within finite regions.

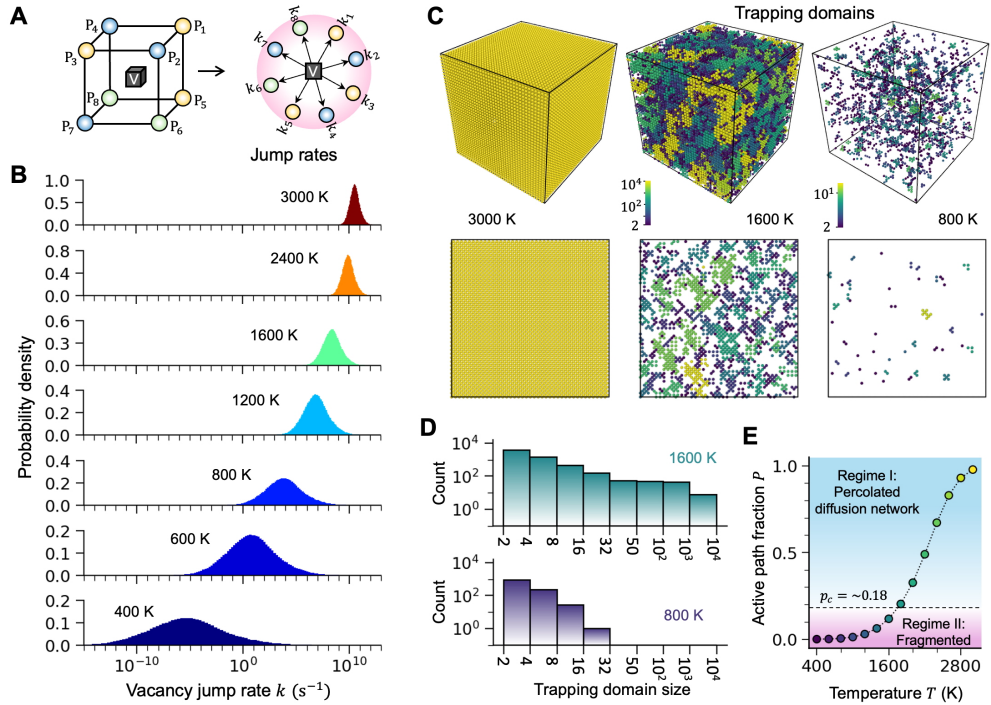

**Fig. 3. Kinetic heterogeneity and the topological breakdown of diffusion network.** (A) Schematic of vacancy migration in bcc lattice. Local chemical disorder defines a unique activation barrier and jump rate $k_i$ for each path. (B) Evolution of vacancy jump-rate probability density distributions from 3000 to 400 K. The distribution transitions from a narrow and nearly uniform peak at high temperatures to an ultra-broad profile spanning more than twelve orders of magnitude at 400 K. (C) Spatial visualization of vacancy trapping domains. At 3000 K, the lattice is fully connected; at 1600 K, the network fragments into discrete clusters; by 800 K, the domains break into nanoscopic cages. (D) Trapping domain size distributions at 1600 and 800 K, illustrating the shift from multi-thousand-site clusters to nano-domains of only a few atomic volumes. (E) Percolation analysis of the active path fraction $P$ as a function of temperature. The diffusion network shows a transition from a percolated (Regime I) to a fragmented and non-percolating state (Regime II).

### Heterogeneous kinetics and diffusion path fragmentation

In elemental W, a vacancy jumps to any of its eight nearest neighboring sites with a uniform rate, $k_i = \nu_0 \exp(-\Delta E/k_B T)$. Each jump path $i$ in WMoTa, however, can possess a unique local chemical environment and hence jump rate $k_i$ (Fig. 3A). We calculate the complete distribution of these barriers and their associated rates for 1,024,000 diffusion paths. The probability distributions of jump rates for temperatures from 3000 K down to 400 K are presented in Figure 3B. At 3000 K, the high thermal energy $k_B T$, obscuring these barrier variations, results in a narrow jump rate distribution and therefore near random walk behavior. At 400 K, the jump rates span over twelve orders of magnitude from $\sim 10$ s$^{-1}$ to $\sim 10^{-12}$ s$^{-1}$, marking significant diffusion heterogeneity where jump becomes highly anisotropic and path-dependent (Fig. 3B).

This kinetic heterogeneity induces a fundamental topological transition in the diffusion path connectivity (Fig. S6). By mapping the spatial distribution of available and accessible paths, there is an emergence of discrete domains (Fig. 3C). While the diffusion paths remain fully connected and percolated at 3000 K, it fragments into isolated clusters at 1600 K. These clusters represent regions of high local mobility surrounded by impenetrable barriers, where the probability of inter-domain hopping is vanishingly low. The size of these trapping domains decreases with temperature (Fig. S7), with the largest cluster occupying $10^4$ atomic volumes at 1600 K and falling drastically to just 32 atomic volumes at 800 K (Fig. 3D). Quantifying the active path fraction reveals a striking percolation transition (Fig. 3E). At lower temperatures, the connectivity of these paths drops rapidly, crossing a critical threshold near 1600 K. Below this threshold (Regime II), the percolated network breaks apart into a collection of isolated traps, interpreting the loss of long-range transport.

**Impeded vacancy cluster growth**

To evaluate how kinetic trapping impacts defect microstructure evolution, we irradiate WMoTa alloys across four orders of magnitude of damage, from 0.0021 to 21 dpa. Bright-field scanning transmission electron microscopy (BF-STEM) reveals the formation of nanometer-scale clusters across all dose levels (Fig. 4A). Remarkably, despite the 10,000-fold increase in irradiation dose, these clusters exhibit negligible growth, maintaining a mean size below 5 nm even at 21 dpa (Fig. 4A). This stands in contrast to pure W, where defect clusters and dislocation loops typically exceed hundreds of nanometers before reaching 1 dpa (*2–4*). In WMoTa, increasing the dose primarily drives defect density rather than size, and small defects are kinetically sequestered within local domains.

In bcc systems, irradiation-induced defects can manifest as prismatic dislocation loops on {111} and {010} habit planes, with Burgers vectors of 1/2⟨111⟩ and ⟨010⟩, respectively. In complex concentrated alloy, however, identifying the vacancy or interstitial nature of these loops becomes even more challenging due to chemical and lattice disorder. Coupling 4D-STEM strain mapping with atomistic simulations, we resolve the specific character of these localized defects and confirm that the suppressed clusters have vacancy-type features.

High-angle annular dark-field (HAADF) imaging along the $[\bar{1}0\bar{1}]$ zone axis resolves defect clusters with edge-on condition of both $(\bar{1}11)$ and $(010)$ planes (Fig. S8-S9). Atomic-resolution images of $(\bar{1}11)$ loop show an elliptical disordered region elongated along the habit plane within the crystalline matrix (Fig. 4B). The associated 4D-STEM strain mapping reveals a diffuse tensile strain field ($\varepsilon_{[\bar{1}11]}$) normal to the habit plane, with nearly vanished strain ($\varepsilon_{[12\bar{1}]}$) parallel to the plane. This strain signature is a clear characteristic of vacancy-type prismatic loop, which produces lattice dilation along its Burgers vector and a minimal strain in the loop plane. Analysis of defect cluster on the (010) plane shows a similar tensile strain signature along its Burgers vector [010], corroborating its vacancy character (Fig. 4C). Compared with the $(\bar{1}11)$ loop, the strain field of (010) loop is more spatially localized with shortened strain tail, consistent with the narrower core width expected for a larger Burgers vector in bcc metals (*26, 27*).

To uncover the atomistic origins of the diffuse strain and distorted core structure, we compare the relaxed defects in W and WMoTa using atomistic modeling for both $(\bar{1}11)$ loop (Fig. 4D) and (010) loop (Fig. S10). In W, the vacancy cluster maintains an ordered configuration, with vacancies located on three adjacent ($\bar{1}11$) planes. The neighboring atoms experience a small off-lattice displacement field, resulting in a concentrated strain field (Fig. 4D). Distinct from W, the loop in WMoTa adopts a markedly perturbed configuration (Fig. 4E). Atomic off-lattice relaxation extends much into the surrounding lattice, accompanied by a disordered core and a diffuse strain field. Vacancies no longer remain confined to three adjacent planes but instead delocalize and spread across multiple planes. These characteristics stem from vacancy solute interactions, which become predominant over ideal structural ordering in W, driving heterogeneous atomic relaxation and giving rise to a rugged, distorted defect structure.

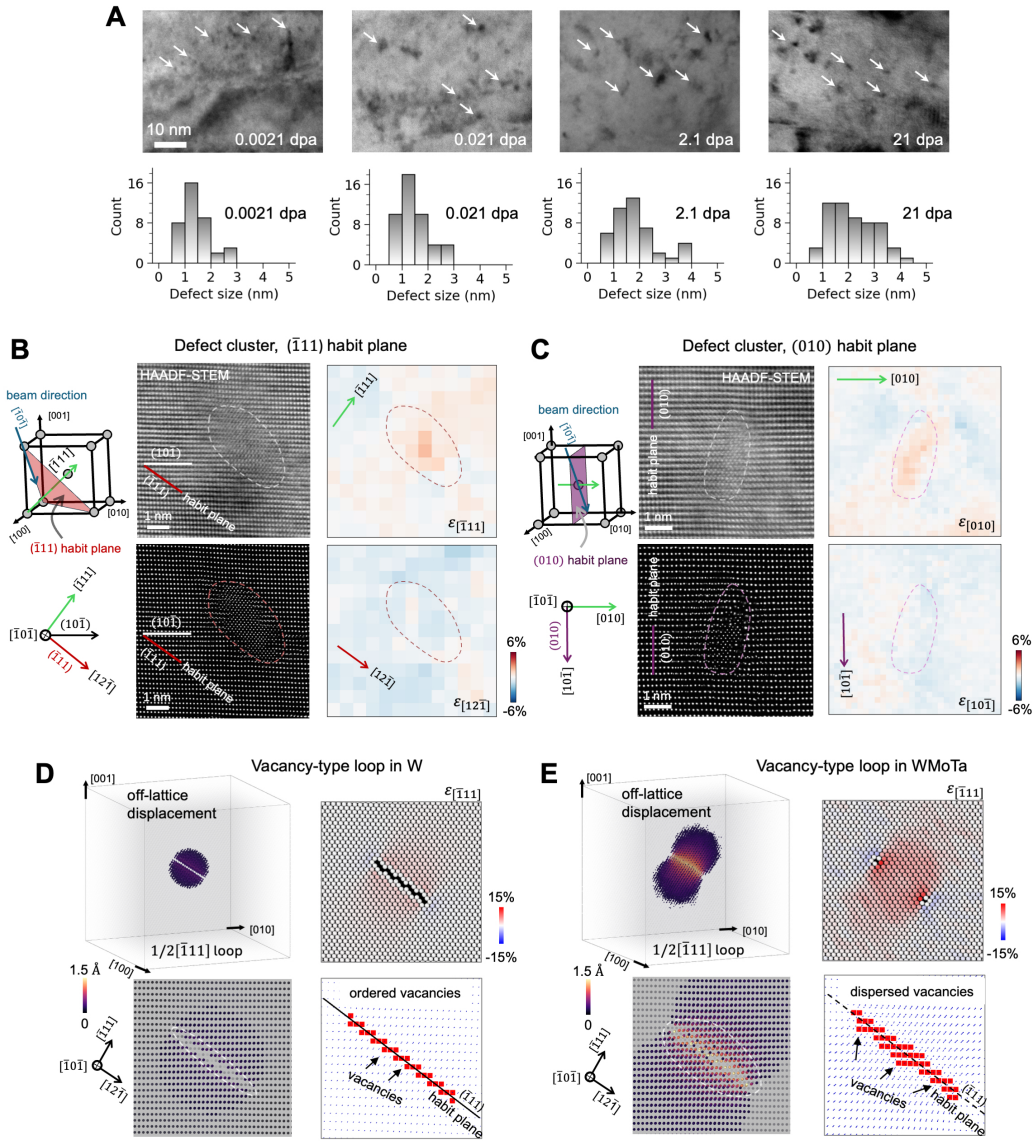

**Fig. 4. Suppression of defect coarsening and vacancy delocalization in irradiated WMoTa.** (A) BF-STEM images and defect size distribution for WMoTa irradiated from 0.0021 to 21 dpa. Clusters remain below ~5 nm even at high damage levels, with increased dose primarily driving defect density. (B) Atomic-resolution HAADF-STEM and 4D-STEM strain mapping of defect cluster on the ($\bar{1}11$) habit plane. The diffuse tensile strain field $\varepsilon_{[\bar{1}11]}$ normal to the plane (along

Burgers vector direction) and minimal in-plane strain $\varepsilon_{[12\bar{1}]}$ identifies the cluster as a vacancy-type defect. (C) HAADF-STEM images and strain analysis of a vacancy-type defect cluster on the (010) plane, showing a shorter strain tail along the plane normal direction [010]. (D) Simulated atomic structure and strain distribution of a $1/2[\bar{1}11]$ vacancy-type loop in W, showing ordered vacancies confined to three adjacent planes and localized off-lattice displacements. (E) Equivalent loop in WMoTa exhibiting extended off-lattice displacement, vacancies dispersed across multiple planes, diffuse strain field, and disordered core resulting from local chemical fluctuations. Scale bars: (A) 10 nm; (B and C) 1 nm.

**Discussion and conclusions**

Since jump rates depend exponentially on activation energy, vacancy jump frequencies in WMoTa can vary over ten orders of magnitude. This creates a kinetically constrained environment, where a subset of rapid diffusion paths is embedded within a largely inactive network. When the fraction of active paths falls below the percolation threshold, vacancies become kinetically confined within isolated segments of the lattice. This fragmentation statistically shields vacancies from coalescing across distant domains, starving clusters of the flux needed for growth. Notably, this trapping is intrinsic to the random solid solution and does not necessitate pre-existing short-range order (SRO). This distinction is important because SRO, stemming from thermodynamic equilibrium, is susceptible to irradiation-induced disordering. Nevertheless, our analysis of aged WMoTa indicates that SRO can further broaden the barrier distribution, intensifying jump selectivity and kinetic trapping (Fig. S13).

Unlike the well-defined prismatic loops in pure metals (*28*), dislocation loops in this alloy exhibit a highly disordered core and an undulating profile, suggesting a potent interaction between solute atoms and vacancy clusters. As a prismatic loop diffuses along its Burgers vector, its motion traverses a spatially varying chemical environment and a hierarchical energy landscape where local basins are nested within large "metabasins" (*29*). The metabasin would exert a consistent back force on the dislocation, retarding its escape and inhibiting the one-dimensional, fast motion commonly occurring in pure metals (*12, 17, 30*). Furthermore, the chemical complexity of WMoTa creates a broad formation enthalpy distribution, allowing individual vacancies to occupy locally favorable, low-energy sites (Fig. S12A). This energetic heterogeneity lowers the free energy of the dispersed-vacancy state relative to a compact cluster, reducing the thermodynamic driving force for aggregation (Fig. S12).

The diffusion network fragmentation and restrained cluster growth identified in WMoTa are not anomalous; rather, they are emergent properties of extreme jump-rate heterogeneity. The FP-NNK model allows these phenomena to be predictively discovered in complex and large compositional spaces. Different from machine learning interatomic potentials (MLIPs) optimized to replicate ab-initio atomic forces and energies (*31*) or many ML models predicting bulk thermodynamics like elastic modulus or melting temperature (*32*), FP-NNK directly targets transition states necessary to resolve slow defect kinetics with ab initio accuracy. In an era where AI can propose millions of stable structures (*33, 34*), representative data and models that explicitly capture activation states and defects remain largely lacking (*35*). By bridging DFT-level precision with experimentally relevant scales, future event-centric AI models will provide a practical route to probe material degradation far beyond the temporal regimes accessible to conventional molecular dynamics.

In summary, the transition from a monolithic W lattice to the WMoTa alloy introduces an extreme kinetic heterogeneity that fundamentally transforms vacancy transport. Coupling predictive modeling with high-dose irradiation experiments, we show that diffusion kinetics can be tailored to delocalize defects and suppress coalescence, enhancing intrinsic radiation resistance. Atomic-scale characterization and 4D-STEM confirm that vacancy cluster growth is kinetically arrested, showing negligible growth even at a raised dose of 21 dpa. This study, revealing jump-rate heterogeneity, diffusion path fragmentation, and their role in governing defect evolution, demonstrates a percolation-engineered paradigm for controlling defect evolution and promoting intrinsic radiation resistance, broadly impacting the long-standing pursuit and endless search for materials that endure the extreme frontiers (*36*) of advanced and efficient energy systems for a sustainable society.

**Acknowledgments**

**Funding:** The work was supported by the U.S. Department of Energy (DOE), Office of Basic Energy Sciences, under Award DE-SC0022295. The authors acknowledge the use of facilities and instrumentation at the UC Irvine Materials Research Institute (IMRI) supported in part by the National Science Foundation Materials Research Science and Engineering Center program through the UC Irvine Center for Complex and Active Materials (DMR-2011967). TJR's effort was supported by the U.S. Department of Energy, Office of Science, Basic Energy Sciences, under Award No. DE-SC0025195. G.T., B.P.U, and P.C. acknowledge support from University of California Office of the President (UCOP), National Laboratory Fees Research Program L26CR10135. Ion irradiation was performed at the Center for Integrated Nanotechnologies, an Office of Science User Facility operated for the U.S. Department of Energy (DOE) Office of Science.

**Author contributions:** P.C. conceived the research idea and advised the project. B.X. implemented the model and performed kinetics simulations. B.J.X. performed TEM characterizations and 4D-STEM strain measurements. W.Z. carried out DFT calculations. E.L. and K.H. performed ion irradiation experiments. E.B. prepared the WMoTa samples. H.C., M.P.S., G.T., T.J.R., J.T., H.H., B.P.U., and K.H., contributed to the discussion of the results. B.X., B.J.X., and P.C. analyzed the data and presented the figures. P.C. wrote the manuscript with information input from B.X. and B.J.X.

**Competing interests:** The authors declare no competing interests.

**Data and materials availability:** All data generated for this study are included in the main text and its supplementary materials. The FP-NNK model, source code, and data are available at the GitHub repository: https://github.com/UCICaoLab/First-principle-NNK.

# Main Figures 1-4

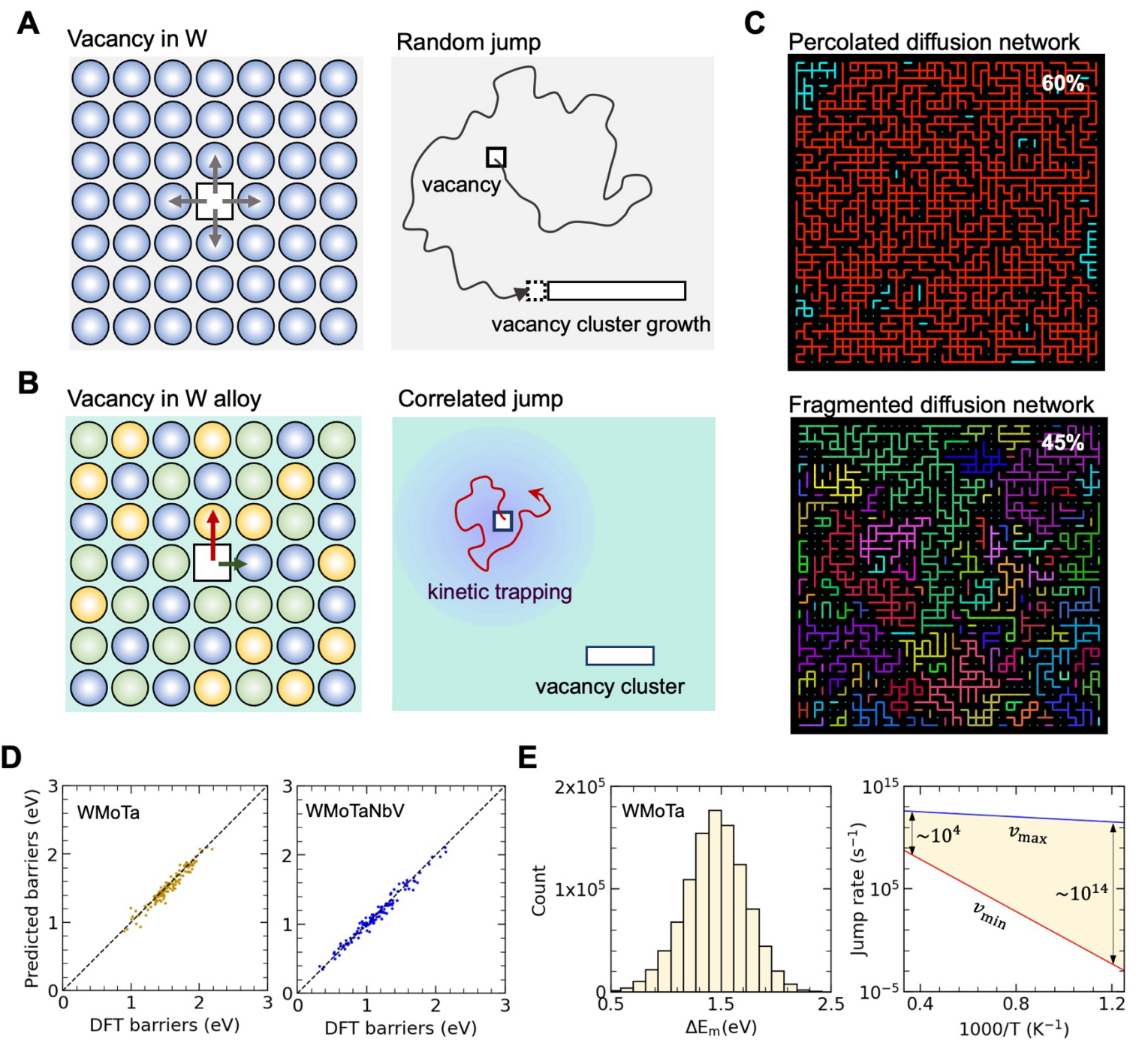

**Fig. 1. Diffusion pathways, network percolation and fragmentation, and diffusion barrier prediction in large compositional spaces.** (A) In pure W, a uniform migration barrier permits vacancies to jump randomly in all crystallographic directions, rendering random walks and long-range diffusion. (B) In alloys, local chemical fluctuations introduce heterogeneous migration barriers, leading to correlated jumps and potential confinement. (C) Conceptual illustration of diffusion network percolation and fragmentation. Above the percolation threshold (top, 60%), accessible pathways form a connected network, and below the threshold (bottom, 45%), the network fragments into isolated domains. (D) FP-NNK predictions of migration barriers compared with DFT values for ternary and quinary W alloys. (E) Statistical distribution of vacancy diffusion barriers in WMoTa spanning 0.5 to 2.3 eV, corresponding to heterogeneous jump rates.

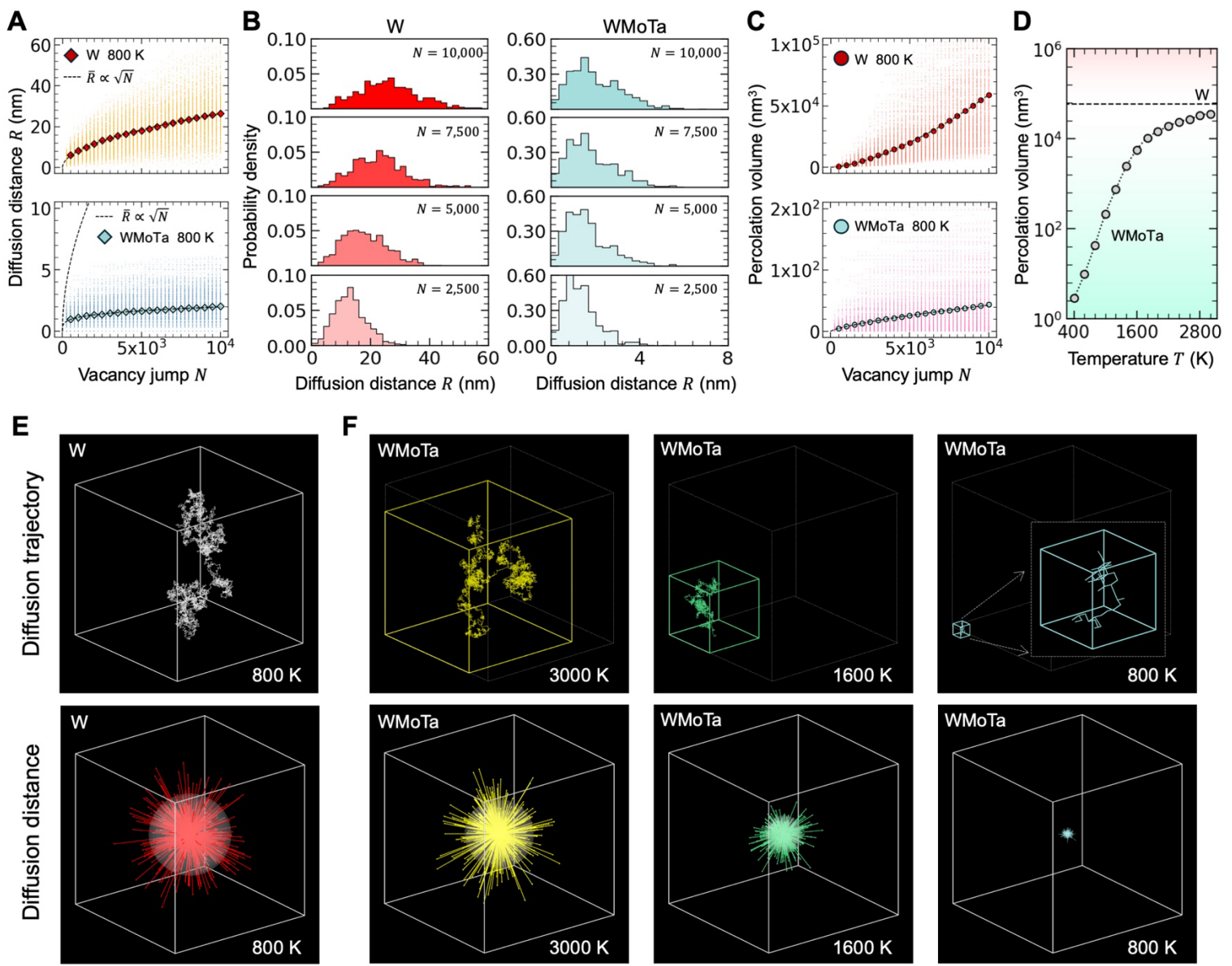

**Fig. 2. Vacancy diffusion distance, trajectories, and kinetic trapping.** (A) Mean diffusion distance $\bar{R}$ as a function of jump number $N$ for W and WMoTa at 800 K. Scattered points represent 500 independent vacancy trajectories, and filled symbols denote the average. The dashed line indicates the classical $\bar{R} \propto \sqrt{N}$ scaling. (B) Probability density distributions of diffusion distance in W (left) and WMoTa (right) after 2,500, 5,000, 7,500, and 10,000 jumps. W exhibits progressively broadening front, whereas WMoTa remains bounded at ~6 nm. (C) Evolution of percolation volume as a function of vacancy jumps. (D) Temperature dependence of the percolation volume, showing a substantial decrease in WMoTa relative to W at 400 K. (E) Representative vacancy trajectory (top) and the diffusion vectors (bottom) for 500 vacancies in W after 10,000 jumps at 800 K. (F) Corresponding trajectories and diffusion vectors in WMoTa after the same number jumps at 3000, 1600, and 800 K.

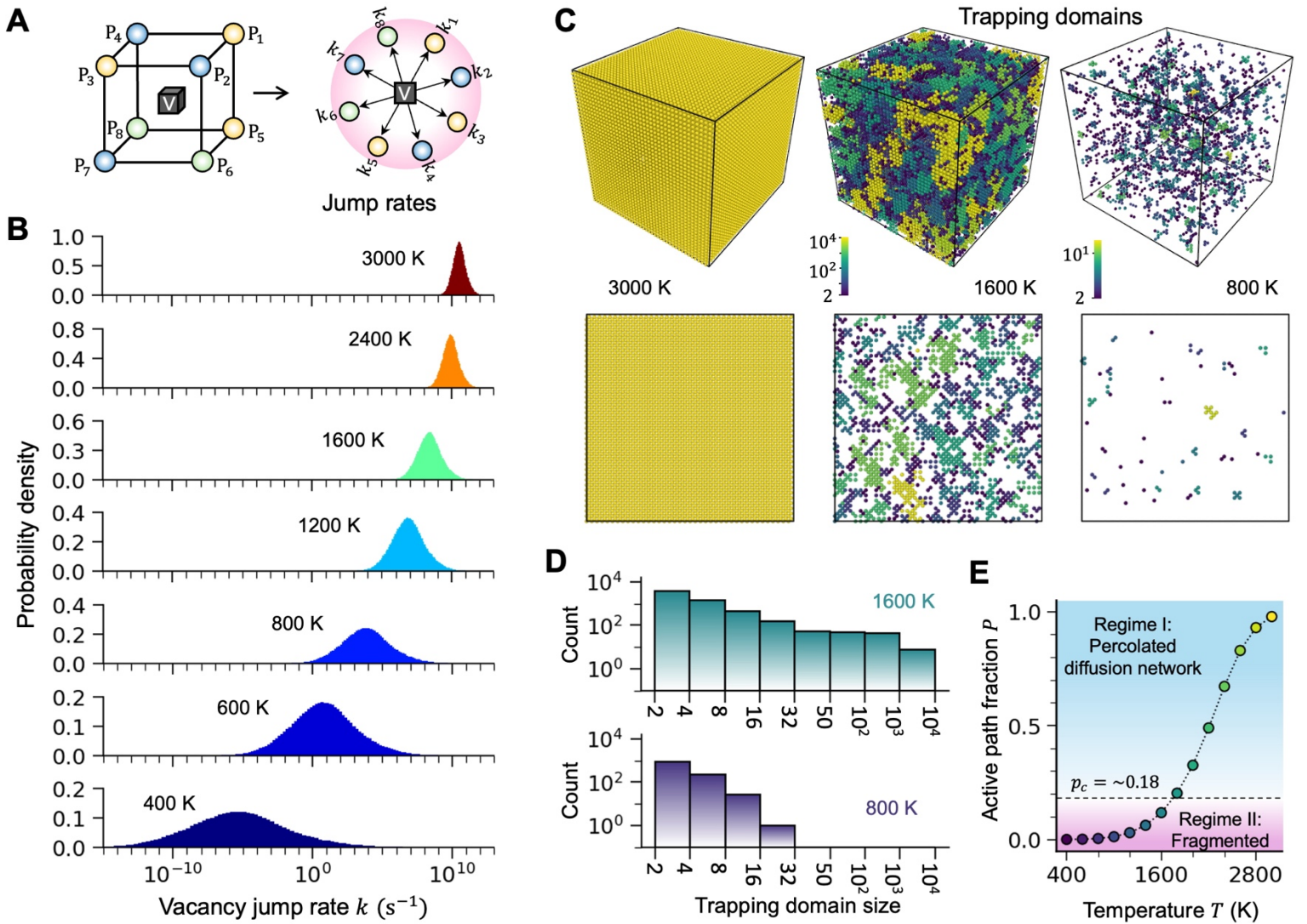

**Fig. 3. Kinetic heterogeneity and the topological breakdown of diffusion network.** (A) Schematic of vacancy migration in bcc lattice. Local chemical disorder defines a unique activation barrier and jump rate $k_i$ for each path. (B) Evolution of vacancy jump-rate probability density distributions from 3000 to 400 K. The distribution transitions from a narrow and nearly uniform peak at high temperatures to an ultra-broad profile spanning more than twelve orders of magnitude at 400 K. (C) Spatial visualization of vacancy trapping domains. At 3000 K, the lattice is fully connected; at 1600 K, the network fragments into discrete clusters; by 800 K, the domains break into nanoscopic cages. (D) Trapping domain size distributions at 1600 and 800 K, illustrating the shift from multi-thousand-site clusters to nano-domains of only a few atomic volumes. (E) Percolation analysis of the active path fraction $P$ as a function of temperature. The diffusion network shows a transition from a percolated (Regime I) to a fragmented and non-percolating state (Regime II).

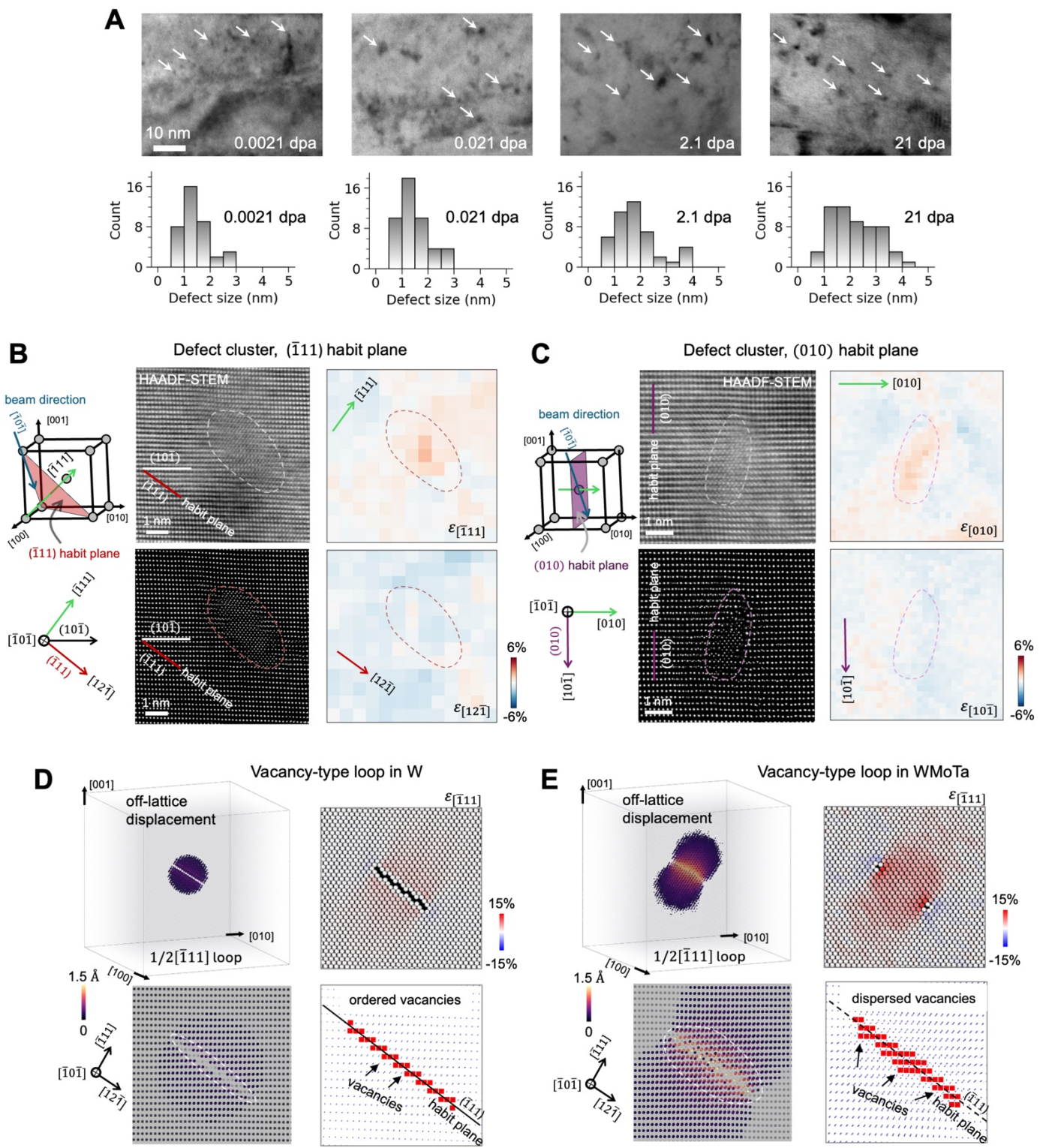

**Fig. 4. Suppression of defect coarsening and vacancy delocalization in irradiated WMoTa.** (A) BF-STEM images and defect size distribution for WMoTa irradiated from 0.0021 to 21 dpa. Clusters remain below ~5 nm even at high damage levels, with increased dose primarily driving defect density. (B) Atomic-resolution HAADF-STEM and 4D-STEM strain mapping of defect cluster on the ($\bar{1}$11) habit plane. The diffuse tensile strain field $\varepsilon_{[\bar{1}11]}$ normal to the plane (along Burgers vector direction) and minimal in-plane strain $\varepsilon_{[12\bar{1}]}$ identifies the cluster as a vacancy-type defect. (C) HAADF-STEM images and strain analysis of a vacancy-type defect cluster on the (010) plane, showing a shorter strain tail along the plane normal direction [010]. (D) Simulated atomic structure and strain distribution of a 1/2[$\bar{1}$11] vacancy-type loop in W, showing ordered vacancies confined to three adjacent planes and localized off-lattice displacements. (E) Equivalent loop in WMoTa exhibiting extended off-lattice displacement, vacancies dispersed across multiple planes, diffuse strain field, disordered core resulting from local chemical fluctuations. Scale bars: (A) 10 nm; (B and C) 1 nm.